\documentclass[a4paper,twocolumn,showpacs,amsmath,amssymb,floatfix,prl,preprintnumbers,footinbib]{revtex4}
\usepackage{graphicx}
\usepackage{dcolumn}
\usepackage{bm}

\begin{document}


\title{Analytic Model for the Energy Spectrum of a Graphene Quantum Dot in a Perpendicular Magnetic Field}
\author{S.~Schnez$^1$, K.~Ensslin$^1$, M.~Sigrist$^2$, and T.~Ihn$^1$}

\affiliation{$^1$Solid State Physics Laboratory, ETH Z\"urich, 8093 Z\"urich, Switzerland\\
$^2$Institute for Theoretical Physics, ETH Z\"urich, 8093 Z\"urich, Switzerland}

\begin{abstract}
We analytically calculate the energy spectrum of a circular graphene quantum dot with radius $R$ subjected to a perpendicular magnetic field $B$ by applying the infinite-mass boundary condition. We can retrieve well-known limits for the cases $R,B\to \infty$ and $B\to 0$. Our model is capable of capturing the essential details of recent experiments. Quantitative agreement between theory and experiment is limited due to the fact that a circular dot deviates from the actual experimental geometry, that disorder plays a significant role, and that interaction effects may be relevant.
\end{abstract}

\maketitle


\section{I. Introduction}
Shortly after its discovery four years ago \cite{Novoselov04}, graphene triggered tremendous research both theoretically and experimentally \cite{Geim07, Neto07}. The excitement was partly created by the fact that charge carriers in graphene are described by the Dirac equation for massless particles and a linear dispersion relation. In particular, there is no energy gap between valence and conduction band. Impressive experimental results like the unconventional quantum Hall effect \cite{Novoselov05, Zhang05} and Klein tunneling \cite{Stander08} were achieved and can be explained with these special properties. First experiments on graphene quantum dots were carried out to study their energy spectra in a perpendicular magnetic field \cite{Ponomarenko08, Schnez08}. The understanding of their spectra, especially how they may differ from a conventional 2D electron system described by the Schr\"odinger equation, is emerging \cite{Chen07, Zhang08}.

Here we look at the energy spectrum obtained by magnetic field spectroscopy in \cite{Schnez08} from the theoretical side. We derive an analytic expression for the energy spectrum in a perpendicular magnetic field in the first part, look at different limits in the second part, and compare theory and experiment in the third part of Sec. II. We conclude with some final remarks in Sec. III.

\section{II. Energy Spectrum in a Perpendicular Magnetic Field}
\subsection{Derivation}
We start from the free Dirac equation expressed in cylindrical coordinates and include a magnetic field oriented normal to the graphene sheet. We use the symmetric gauge for the vector potential, $\bm{A} = B/2(-y, x,0)=B/2(-r\sin\phi,r\cos\phi,0)$ with $\phi$ being the polar angle. The Hamiltonian then reads
\begin{equation}\label{Dirac_Hamiltonian}
  H = v_F\left(\bm{p} +e\bm{A}\right)\cdot \bm{\sigma} + \tau V(r)\sigma_z,
\end{equation}
and the Dirac equation is $H\psi(r,\phi)=E\psi(r,\phi)$ with the wave function being a two-component spinor, $\psi(r,\phi)=\left(\psi_1(r,\phi),\psi_2(r,\phi)\right)$. The charge of an electron is given by $-e$, $v_F$ is the Fermi velocity, and ${\bm \sigma}=(\sigma_x,\sigma_y)$ are Pauli's spin matrices in the basis of the two sublattices of A- and B-atoms. The electron spin is neglected in our analysis. A mass-related potential energy $V(r)$ is coupled to the Hamiltonian via the $\sigma_z$ Pauli matrix. 
The mass in the dot is zero, $V(r)=0$ for $r<R$, but tends to infinity at the edge of the dot, $V(R)\to \infty$. In this way, charge carriers are confined inside the quantum dot. This leads to the infinite-mass boundary which yields the simple condition that $\psi_2/\psi_1 = \tau i \exp[i\phi]$ for circular confinement \cite{Berry87}. We are aware that the boundary condition \cite{Akhmerov08} can be different in the experimental situation. The parameter $\tau$ takes the two values $\pm 1$ which leads to a distinction between the two valleys K and K' described by the Dirac equations of Eq. \eqref{Dirac_Hamiltonian}. 
Hence, in the following we can set $V(r)=0$ and we will respect the different energy spectra of the K- and K'-valleys via the boundary condition. 

Since the operator for the total angular momentum, $J_z = l_z+\frac{\hbar}{2}\sigma_z$ commutes with $H$, $ \left[H,J_z\right] = 0$, we can construct simultaneous eigenspinors for $H$ and $J_z$ ($m$ being an integer),
\begin{equation}
  \psi(r,\phi) =  e^{im\phi}\left(\begin{array}{c}\chi_1(r)\\e^{i\phi}\chi_2(r)\end{array}\right).
\end{equation}
Plugging this expression into the Dirac equation and decoupling the system of differential equations, we arrive at a second-order differential equation for, e.~g., $\chi_1(r)$ which depends only on $r$, 
\begin{equation}
  \left[\partial_r^2+\frac{1}{r}\partial_r - \frac{m+1}{l_B^2} - \frac{m^2}{r^2}- \frac{r^2}{4l_B^4}+k^2\right]\chi_1(r) = 0.
\end{equation}
The energy $E$ is related to the wave vector $k$ according to $E=\hbar v_F k$. We have introduced the magnetic length $l_B=\sqrt{\hbar/(eB)}$. In order to solve this differential equation, we make the {\it ansatz} $\chi_1(r)=r^m \exp\left[-r^2/4l_B^2\right] \xi\left(r^2\right)$. This yields the associated Laguerre differential equation
\begin{equation}\label{ass_Laguerre_diff_eq}
  \left[\tilde{r} \partial_{\tilde{r}}^2 + \left(m+1 - \frac{\tilde{r}}{2l_B^2}\right)\partial_{\tilde{r}}+ \frac{k^2l_B^2 - 2(m+1)}{4l_B^2}\right] \xi\left(\tilde{r}\right)=0.
\end{equation}
with $\tilde{r}:=r^2$. The solution is $\xi\left(\tilde{r}\right) = c\, L\left(k^2l_B^2/2-(m+1),m,\tilde{r}/2l_B^2\right)$, where $L(a,b,x)$ is the generalized Laguerre polynomial and $c$ is a normalization constant. The second linearly independent solution of Eq. \eqref{ass_Laguerre_diff_eq}, the confluent hypergeometric function of the second kind, does not appear in the wave function $\xi$ because it cannot be normalized. With the final result for $\xi$ and hence for $\psi_1$, $\psi_2$ can be derived directly from Eq. \eqref{Dirac_Hamiltonian}. The wave functions finally read
\begin{widetext}
  \begin{equation}
    \begin{split}
      \psi_1(r,\phi) &= c\, e^{im\phi} r^m e^{-r^2/4l_B^2}L\left(\frac{k^2l_B^2}{2}-(m+1),m,r^2/2l_B^2\right),\\
      \psi_2(r,\phi) &= c\, ie^{i(m+1)\phi}r^m e^{-r^2/4l_B^2}\frac{r/l_B}{kl_B} \left[L\left(\frac{k^2l_B^2}{2}-(m+2),m+1,r^2/2l_B^2\right)+L\left(\frac{k^2l_B^2}{2}-(m+1),m,r^2/2l_B^2\right)\right].
    \end{split}
  \end{equation}
Employing the boundary condition discussed above, we finally get an implicit equation for determining $k$, namely
\begin{equation}\label{energy_equation}
  \left(1-\tau \frac{kl_B}{R/l_B} \right) L\left(\frac{k^2l_B^2}{2}-(m+1),m,\frac{R^2}{2l_B^2}\right) + L\left(\frac{k^2l_B^2}{2}-(m+2),m+1,\frac{R^2}{2l_B^2}\right) =0.
\end{equation}
\end{widetext}
Since the generalized Laguerre polynomials are oscillatory functions, there is an infinite number of $k_n$'s for given $B$, $m$, and $\tau$ which fulfill the above equation. This defines the radial quantum number $n$ which labels the roots of the left part of Eq. \eqref{energy_equation}. Here, we restrict ourselves to positive solutions for $k$ so that $k_1\geq 0$. Therefore, the energy spectrum $E(n,m,\tau)$ of electrons confined to a circular graphene quantum dot which is exposed to a perpendicular magnetic field is determined through Eq. \eqref{energy_equation}. The relation $-E(n,m,\tau) = E(n,m,-\tau)$ is a manifestation of the electron-hole symmetry.

\subsection{Limits for $B\to 0$ and $R/l_B\to\infty$}
For a better understanding of Eq. \eqref{energy_equation}, we will look at the two limits $B\to 0$ and $R/l_B\to\infty$ separately. Bessel functions of the first kind can be expressed as a limit of the generalized Laguerre polynomial \cite{Abramowitz},
\begin{equation}
  \lim_{a\to\infty} \left[\frac{1}{a^b} L\left(a,b,\frac{x}{a}\right)\right] = x^{-b/2}J_b\left(2\sqrt{x}\right).
\end{equation}
Using this property, Eq. (\ref{energy_equation}) can be simplified to
\begin{equation}\label{B_to_0}
  \tau J_m(kR) = J_{m+1}(kR)
\end{equation}
in the limit $B\to 0$. This is the result already derived in \cite{Berry87}. The relation can be used to estimate the number of charge carriers confined on a graphene dot when the energy of an excited state is measured \cite{Ponomarenko08, Schnez08}. Moreover, we can deduce that $E(n,m,\tau) = E(n,-m-1,-\tau)$ for $B=0$. This is derived from the property $J_m(x) = (-1)^mJ_{-m}(x)$. Hence, pairs of states are degenerate at zero magnetic field. There is no state at zero magnetic field and zero energy \cite{Berry87}. This leads to an energy gap between negative and positive energy states. The size of the energy gap will be discussed below.

Landau levels should be retrieved from Eq. \eqref{energy_equation} if the confinement is lifted. Mathematically, this is achieved for $R/l_B\to\infty$. We express the generalized Laguerre polynomial in terms of the confluent hypergeometric function of the first kind $M(\alpha,\beta,\gamma)$ \cite{Abramowitz}
\begin{equation}
  L(a,b,x)=\left(\begin{array}{c} a+b\\ a\end{array}\right)M(-a,b+1,x).
\end{equation}
For $R/l_B\to \infty$, a power expansion of $M$ is possible and yields to first order \cite{Abramowitz}
\begin{equation}
  M(\alpha,\beta,\gamma) = \frac{\Gamma(\beta)}{\Gamma(\alpha)}e^\gamma \gamma^{\alpha-\beta}\left(1+O\left(|\gamma|^{-1}\right)\right).
\end{equation}
Rewriting the binomial coefficients with Gamma functions $\Gamma(x)$ and using one of their defining relations, $\Gamma(x+1) = x\Gamma(x)$, algebraic manipulations of Eq. \eqref{energy_equation} give
\begin{equation}\label{LL}
    E_m = \hbar v_F k_m = \pm v_F\sqrt{2e\hbar B(m+1)}.
\end{equation}
Hence, we retrieve the well-known Landau levels for graphene. Therefore, there will be a transition, governed by the parameter $R/l_B$, from a regime where the energies of the electrons are dominated by confinement (Eq. \eqref{B_to_0}) to Landau levels (Eq. \eqref{LL}). This transition including these two limiting cases is described by Eq. \eqref{energy_equation}.

\subsection{Numerical Results and Comparison to Experiment}
\begin{figure}
  \includegraphics[width=\linewidth]{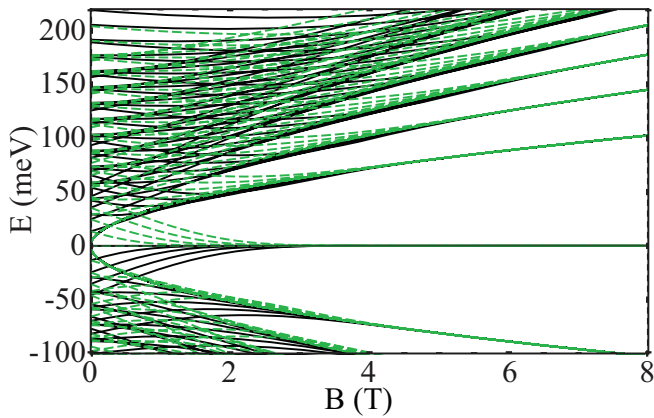}
  \caption{(Color online) Energy spectrum of a quantum dot with $R = 70\,\textrm{nm}$. The formation of the lowest Landau levels can be seen as predicted by Eq. \eqref{LL}. Energy states for $\tau=+1$ are drawn with black, solid lines, those for $\tau=-1$ with green, dashed lines.\label{fig1}}
\end{figure}

We evaluate Eq. \eqref{energy_equation} for a dot of radius $R = 70\,\textrm{nm}$ which is about the same size as the device measured in \cite{Schnez08}. The energy spectrum as a function of magnetic field is shown in Fig. \ref{fig1} for $m=-4,\dots, 4$ and $n=1,\dots, 6$. For $B=0$, the energy states are not equidistant. For higher magnetic fields, we can see the formation of Landau levels as it is expected according to the previous discussion. The zero energy Landau level is formed by states with quantum number $\tau=-1$ and $E>0$ and those with $\tau=+1$ and $E<0$. For completeness, we plot the first negative energy states as well in Fig. \ref{fig1}.

The lowest positive energy state has an energy of about $4\,\textrm{meV}$ for $B=0$. This gives an energy gap of around $8\,\textrm{meV}$ between electron and hole states. Since the energy gap to the next excited state is much lower, the electron-hole transition may be detected experimentally by a confinement enhanced energy.

For quantum dots in semiconductors, the Darwin-Fock model is often used to qualitatively explain the experimental observations.  In contrast to the model presented in this paper, the Darwin-Fock model is based on harmonic confinement giving rise to equidistant and highly degenerate energy levels at $B=0$. However, the general evolution from single-particle states at $B=0$ to Landau levels at high magnetic fields is similar in the two scenarios.

\begin{figure}
  \includegraphics[width=\linewidth]{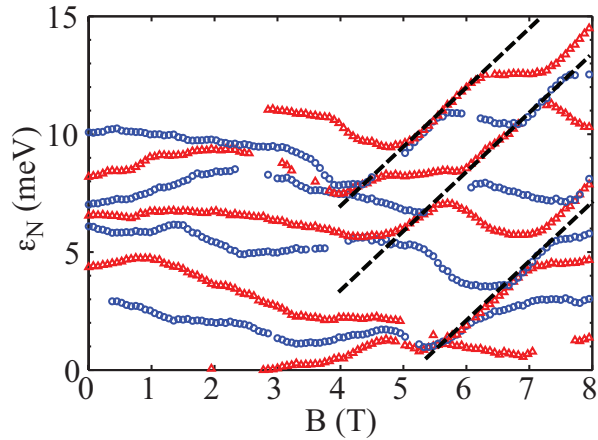}
  \caption{(Color online) Experimental data of a quantum dot with $R = 70\,\textrm{nm}$. This figure is taken from \cite{Schnez08}. The single-particle energy of nine consecutive states (labeled with red triangles and blue circles, respectivelt) is shown. The characteristic slopes of the dashed lines are about $\pm 2.5\,\textrm{meV/T}$.\label{fig2}}
\end{figure}

We recently performed transport spectroscopy measurements on Coulomb blockade resonances in a graphene dot with a radius of about $70\,\textrm{nm}$ \cite{Schnez08}; Fig. \ref{fig2} is the main result of that experiment. It shows the position of conductance resonances as a function of magnetic field. The vertical energy axis was obtained by converting plunger gate voltage into energy using the measured lever arm. Employing the constant-interaction model, the ground state energy of an $N$-particle quantum dot can be written as 
\begin{equation}
  E_{\textrm{gs}}(N) = \sum_{i=1}^N\varepsilon_i(B) + \frac{e^2N^2}{2C_\Sigma} - eN\alpha V_g.
\end{equation}
The single-particle energy $\varepsilon_i(B)$ of the $i$-th particle is given by $E(n,m,\tau)$. The second summand is the electrostatic contribution to the energy with the total capacitance $C_\Sigma$ of the experimental device which we assume to be independent of $B$. The ground state energy can be tuned by a gate voltage $V_g$. The lever arm $\alpha$ is deduced from Coulomb diamond measurements. The experiment described in \cite{Schnez08} was done in the zero-bias regime; in other words we measured the chemical potential $\mu_N = E_{\textrm{gs}}(N) - E_{\textrm{gs}}(N-1)$ of the $N$-th Coulomb resonance. Hence, the single-particle energy $\varepsilon_N(B)$ can experimentally be determined by $\varepsilon_N(B) = e\alpha V_g^\textrm{res}(N,B) + \textrm{const.}$ In Fig. \ref{fig2}, the constant part is subtracted so that consecutive Coulomb resonances (labeled with red triangles and blue circles, respectively) touch each other in one point. Since there is no well-defined zero energy (the electron-hole transition could not be determined in the experiment), an arbitrary offset was subtracted.

\begin{figure}
  \includegraphics[width=\linewidth]{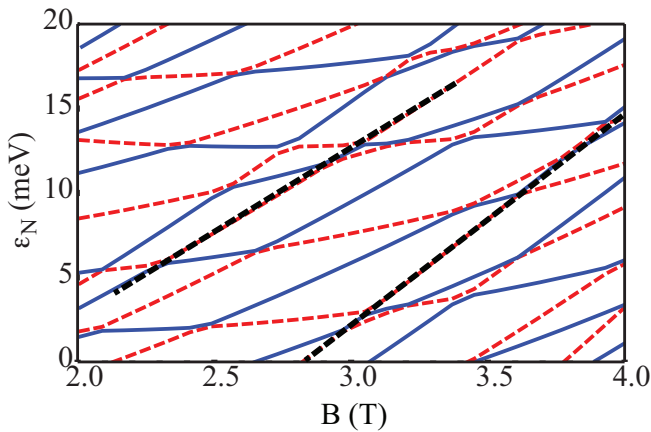}
  \caption{(Color online) A close-up of the single-particle spectrum shown in \ref{fig1}. For comparison to the experimental data, states with a fixed number of particles are consecutively denoted by red, dashed and blue, solid lines. An arbitrary offset is subtracted as described in the text. Slopes marked by the two black, dashed lines vary between $7\,\textrm{meV/T}$ and $12\,\textrm{meV/T}$.\label{fig3}}
\end{figure}

We compare the experimental data to our theoretical model. For this purpose, we zoom into a particular region of Fig. \ref{fig1}. Since we are now interested in states with a constant number of particles, these are -- in analogy to Fig. \ref{fig2} -- shown consecutively with red dashed and blue solid lines in Fig. \ref{fig3}.  The kinks in the energy spectrum for a constant number $N$ of particles occur whenever the $N$-th particle changes its quantum state to stay in the lowest possible single-particle state. In Fig. \ref{fig3}, the slopes are constant and vary between $7\,\textrm{meV/T}$ and $12\,\textrm{meV/T}$. This is in reasonable agreement with our experimental data where we measured slopes of $2.5\,\textrm{meV/T}$. The following three reasons might limit the quantitative agreement between experiment and theory. i) The circular shape is a simplification of the experimental device. ii) Disorder is expected to play a significant role in graphene nanostructures. This is not included in the theoretical model. iii) Since the experiments were not performed in the single-electron regime, interactions effects should be included in a thorough theoretical analysis. Numerical tight-binding calculations or quasi-classical simulations are capable of implementing these aspects \cite{Libisch08}.

\section{III. Summary}
In this paper, we derived an analytic expression for the energy spectrum of a circular graphene quantum dot which is exposed to a perpendicular magnetic field. The boundary condition we employed is the infinite-mass boundary introduced in \cite{Berry87}. This straightforward model is in good qualitative agreement with our recent experiments \cite{Schnez08}. We discuss possible limitations of the model.  

The validity of the infinite-mass boundary condition for graphene is discussed in recent papers \cite{Wurm08}. One needs to evaluate the consequences of different boundary conditions for energy spectra of confined graphene quantum structures. However, apart from the reasonable agreement between theory and experiment, instructive models like the one presented here can be solved analytically with the infinite-mass boundary and give an intuitive understanding of the physics behind such systems.

\emph{Acknowledgement:} Financial support by ETH Z\"urich and the Swiss Science Foundation is gratefully acknowledged.

\emph{Note added:} During preparation of this manuscript, we became aware of similar and more extensive calculations by P. Recher \emph{et al.} \cite{Recher08}.

\end{document}